\documentclass{desyproc}
\usepackage{graphicx}

\begin{document}
\title{Anisotropic flow from hard partons\\ in ultra-relativistic nuclear collisions}

\author{{\slshape Boris Tom\'a\v{s}ik$^{1,2}$, Martin Schulc$^2$}\\[1ex]
$^1$Univerzita Mateja Bela, Tajovsk\'eho 40, 97401 Bansk\'a Bystrica, Slovakia\\
$^2$Czech Technical University in Prague, FNSPE, B\v{r}ehov\'a 7, 11519 Prague 1, 
Czech Republic }

\contribID{96}

\confID{8648}  
\desyproc{DESY-PROC-2014-04}
\acronym{PANIC14} 
\doi  

\maketitle

\begin{abstract}
Anisotropies of hadronic distribution in nuclear collisions are used 
for  determination of properties of the nuclear matter. At the 
LHC it is important to account for the contribution to the flow due to momentum transferred 
from hard partons to the quark-gluon plasma. 
\end{abstract}



In ultrarelativistic nuclear collisions, hadron momentum distributions are azimuthally 
ani\-so\-tro\-pic and 
parametrised with the help of 
Fourier expansion with amplitudes of individual modes usually denoted as $v_n$'s. 
If spectra are summed over a large number of events, symmetry 
constraints dictate all odd amplitudes  to vanish. 
In individual events, however, these constraints are not realised, because 
the event shapes fluctuate. 

In general, the mechanism behind the modification of hadronic spectra is the blue-shift. 
Transverse expansion of the fireball enhances production of hadrons with
higher $p_t$. If the normalisation
and the slope of  $p_t$ spectrum depend on azimuthal angle, this indicates different
transverse expansion velocity in different directions. Expansion is caused by
pressure gradients in the initial state. We thus have a link 
between the initial state of the fireball and the observed hadronic spectra. 
(In fact, here we propose a mechanism which can break this link.)

The link is described by relativistic hydrodynamics.  The scheme is based on 
fundamental conservation laws complemented by the equation of state. In non-ideal hydrodynamics
it also involves transport coefficients, e.g.\ shear and bulk viscosity.  The goal is
to tune them so that hydrodynamic modelling yields results in accord with the observations. 

Unfortunately, there are some problems. The initial conditions 
are unknown. They are set by energy depositions in early partonic interactions. 
Various models predict energy and momentum density profiles
with different levels of spikiness. One can get the same result on flow 
anisotropies with different initial conditions  if one re-tunes the transport 
coefficients \cite{Luzum:2008cw}. This hinders
the determination of viscosities from comparisons to data. The extracted 
values would depend on the assumptions that are made about \emph{unknown} initial 
conditions. 

This problem might be settled  with the help of flow anisotropy 
fluctuations \cite{Heinz:2013wva}. 
Simulations indicate that the values of $v_n$'s in individual events follow to large extent the 
corresponding spatial anisotropies of the initial state \cite{Gale:2012rq,Niemi:2012aj}. The 
departure from this proportionality has also been studied 
\cite{Niemi:2012aj,Floerchinger:2013hza}. The mechanism proposed in the present paper
would break this proportionality since it produces flow 
anisotropy \emph{during} the hydrodynamic evolution \emph{without} the need for 
any anisotropy in the initial state. 



We point out \cite{Tomasik:2011xn}
that in nuclear collisions at  LHC energies
there is more than one dijet pair per event. (We might have to lower the threshold 
for what we count as hard parton; here we use $p_t>3$~GeV/$c$.) They deposit 
most---if not all---of their energy and momentum into the plasma and are fully 
quenched. Since momentum must be conserved, the wakes behind the partons must stream and
carry it. 
Such streams would generate anisotropy of collective expansion in every 
individual event. This leads to elliptic anisotropy even after a summation 
over large number of 
events. Indeed, isotropically produced jets 
generate elliptic anisotropy.  
The important detail is the possibility that the induced 
streams can interact and merge.

Suppose that 
two dijet pairs are produced in a non-central collision. 
The elliptic flow due to 
\begin{wrapfigure}{r}{0.47\textwidth}
\includegraphics[width=0.23\textwidth]{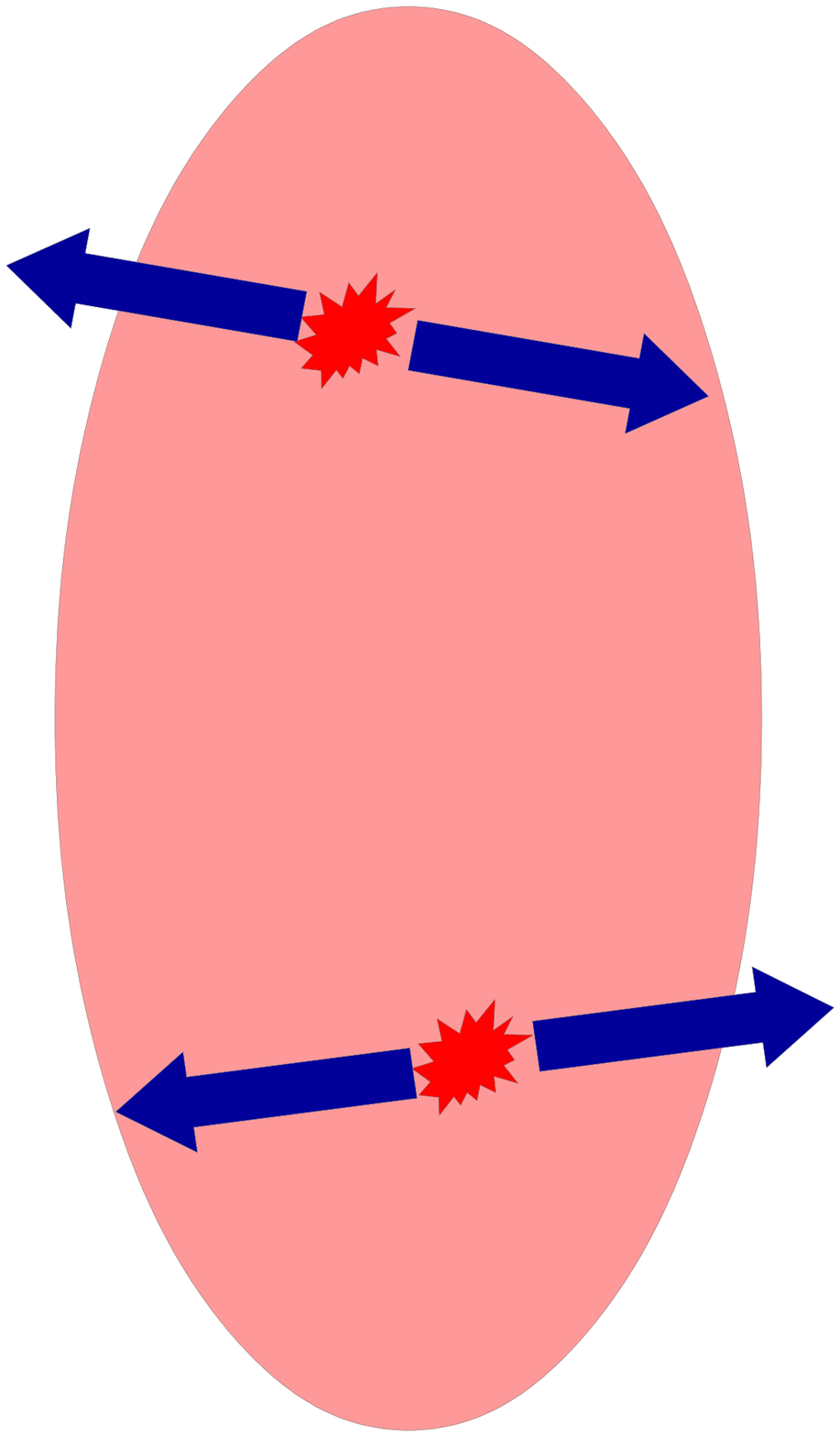}
\includegraphics[width=0.23\textwidth]{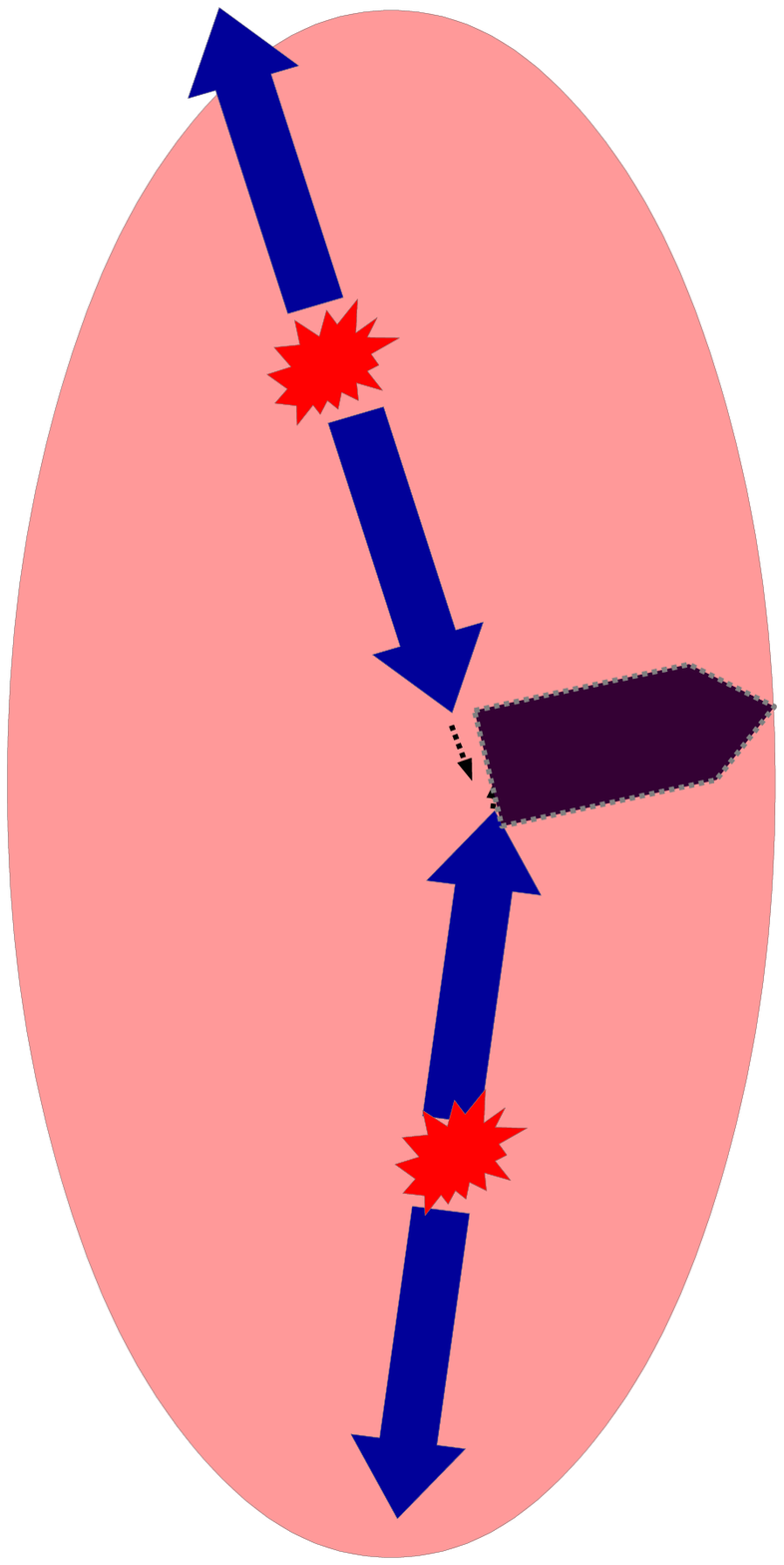}
\caption{\label{f:cartoon}%
Transverse cross-section through the fireball with two dijet pairs produced. Reaction plane 
is horizontal.
Left: two dijets both emitted in the direction of the reaction plane both contribute 
positively to the elliptic flow, which is dominant in the same direction. Right: if 
hard partons are produced off the reaction plane, some of their streams can come together 
and merge. 
}
\end{wrapfigure}
spatial deformation is directed parallel to the reaction plane.
If both pairs are aligned with
this plane, then all streams contribute to positive $v_2$, see
Fig.~\ref{f:cartoon} (left). 
On the other hand, if the jets are oriented under large angle with respect to the reaction 
plane, then the two streams directed inwards can meet, merge into one, and continue in a 
direction given by the sum of their two momenta. They do not contribute to the collective 
flow in their original direction. The chance 
of merger is higher in the latter case than in the former one since there the jets pass each 
other within a narrower path. Perpendicularly to the reaction plane the fireball 
is wider so the streams parallel to the reaction plane
can well proceed without bothering each other. In addition 
to this mechanism, Fig.~\ref{f:cartoon} (right) also suggests that contribution to 
triangular flow is created by the merger of two streams. 


This picture is supported by our simulations.
We developed 3+1D ideal hydrodynamic simulation code 
\cite{Schulc:2013kra,Schulc:2014jma} using the 
SHASTA scheme to handle shocks. We include force term $J^\mu$
\begin{equation}
\label{eq:hydro-f}
\partial_\mu T^{\mu\nu} = J^\nu
\end{equation}
which represents the dragging of the fluid by hard partons \cite{Betz:2008ka}
\begin{equation}
J^\nu = -\sum_i \frac{1}{(2\, \pi\, \sigma_i^2)^{\frac{3}{2}}} \, \exp \left (
- \frac{\left ( \vec x - \vec x_{\mathrm{jet},i} \right )^2 }{2\, \sigma_i^2} \right )\, 
\left ( \frac{dE_i}{dt}, \frac{d\vec P_i}{dt} \right )
\end{equation}
where the sum goes through all hard partons in the system and the width $\sigma_i$ was 
set to 0.3. 

We first checked that indeed the streams are induced behind the partons and that they 
flow even after the partons are fully quenched 
(as was also observed in \cite{Betz:2008ka}). In a 
simulation with  static medium we could see that the streams merge when they meet. 
Then, until their energy is spread over a larger volume, they continue flowing in 
common direction \cite{Schulc:2013kra}. 

The mechanism has been included into more realistic simulation of nuclear collisions. 
In these studies it was not our aim to reach the complete description of data. We rather wanted 
to gain realistic estimate of the influence of our mechanism on the observed anisotropies. 
Therefore, we started our simulations always with smooth initial conditions calculated within 
the optical Glauber model. Any fluctuation on top of non-zero event-averaged flow 
harmonics is then clearly a consequence of hard partons inducing flow anisotropies. 
We start our simulation with uniform profile in longitudinal rapidity stretched over
10 units and cut by half-Gaussian tails at both ends. This feature represents the approximate 
boost-invariance at highest LHC energies. Note that the use of 3+1D hydrodynamic model, 
which makes our simulation distinct from those reported in 
\cite{Andrade:2014swa,Floerchinger:2014yqa}, is
important because the hard partons injected into plasma break the boost invariance 
and thus the possibility to reduce the dimensionality of the hydrodynamic model. 

At the beginning of each event simulation we generate the positions and directions 
of the hard parton pairs. Their number fluctuates according to Poissonian and their 
$p_T$'s follow from \cite{Tomasik:2011xn}
\begin{equation}
E \frac{d\sigma_{NN}}{d^3p} = \frac{1}{2\pi}\, \frac{1}{p_T}\, \frac{d\sigma_{NN}}{dp_T\, dy}
= \frac{B}{(1+p_T/p_0)^n}
\end{equation}
with $B = 14.7$~mb/GeV$^2$,  $p_0 = 6$~GeV and $n = 9.5$. Momenta in a pair are 
back-to-back. The initial positions are generated 
from the distribution of the binary collisions calculated within optical Glauber model. 

In an expanding fireball we assume that the energy loss of a parton scales with the entropy 
density as 
$
{dE}/{dx} = \left . {dE}/{dx}\right |_0 ({s}/{s_0}) 
$
where $s_0=78.2$/fm$^{3}$ (corresponds to energy density 20~GeV/fm$^3$). 
Hydrodynamic description of the collision is finished at the freeze-out hypersurface specified 
by  temperature 150~MeV. Generation of final state hadrons is done with the 
help of THERMINATOR2 \cite{Chojnacki:2011hb} Monte Carlo model. 

\begin{figure}[t]
\includegraphics[width=0.98\textwidth]{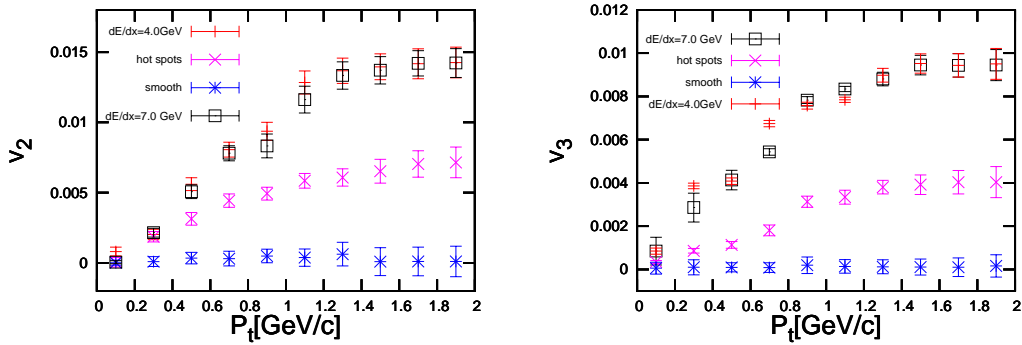}
\caption{%
Anisotropy coefficients from central collisions. Two simulations with hard partons
with different energy loss. One simulation with only energy and no momentum deposition 
(hot spots). One simulation with smooth initial conditions. 
}%
\label{f:central}
\end{figure}

In Fig.~\ref{f:central} we show the $v_n$'s calculated in central collisions. 
To study the effect of momentum
deposition
we simulated 100 evolutions for every setting 
and generated 5 THERMINATOR2 events for each of them. For the momentum loss we made 
simulations with $dE/dx|_0$ set to 4~GeV/fm and 7~GeV/fm and they lead to the same 
momentum anisotropies. Their magnitude indicates that the effect is important
and should be included in realistic simulations. Finally, we also simulated events 
where we put in hot spots with the same energy on top of the smooth initial 
conditions
instead of hard partons. 
They deposit only energy and no momentum, 
and the generated 
flow anisotropies are about half of those initiated by hard partons.

\begin{wrapfigure}{r}{0.51\textwidth}
\includegraphics[width=0.5\textwidth]{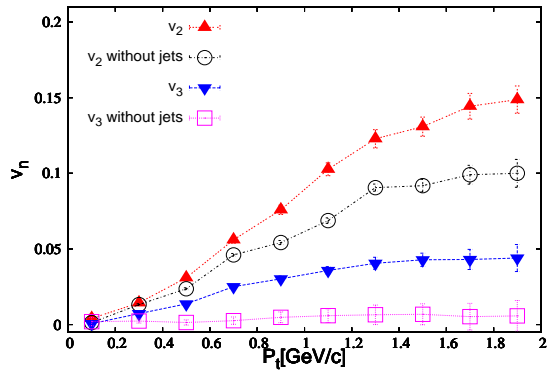}
\caption{\label{f:noncen}%
Coefficients $v_2$ and $v_3$ from 30--40\% centrality events. Results of simulations 
with hard partons are compared with results from smooth initial conditions. 
}
\end{wrapfigure}
Simulations of non-central collisions clearly show that the contribution enhances 
the observed a\-ni\-so\-tro\-pies. In Fig.~\ref{f:noncen} we see about 50\% addition 
to $v_2$ as compared to the case with smooth initial conditions. Triangular anisotropy 
is absent in the initial conditions and thus any $v_3$ is exclusively due to hard partons.


The presented results clearly demonstrate the necessity to include this mechanism 
into realistic hydrodynamic simulations which aim at extracting the properties of quark
matter. For the alignment of the studied effect with the geometry of the fireball
it is crucial to include more than one dijet pair into the simulation, unlike 
done in \cite{Tachibana:2014lja}. The interplay of many generated streams appears important. 

\paragraph{Acknowledgement}
This work was supported in parts by APVV-0050-11, VEGA 1/0457/12 (Slovakia) and 
M\v{S}MT grant  LG13031 (Czech Republic).

\end{document}